\documentclass[11pt]{amsart}
\usepackage{epsfig}
\usepackage{graphicx,color}
\usepackage{amsmath, amssymb,mathabx,mathrsfs}
\usepackage[toc,page]{appendix}
\usepackage{comment}
\usepackage{hyperref}
\usepackage{float}
\usepackage{xcolor}
\usepackage{amsthm}
\usepackage{amsfonts} 
\usepackage{orcidlink}
\newcommand{\authororcid}[1]{%
  \kern0.12em\raisebox{0.6ex}{\orcidlink{#1}}%
}
\usepackage{subcaption}

\usepackage{placeins}
\theoremstyle{definition}

\theoremstyle{remark}

\numberwithin{equation}{section}

\begin{document}
\title[Null-Validated Topological Signatures of Financial Market Dynamics]{Null-Validated Topological Signatures of Financial Market Dynamics}

\author[S. W. Akingbade]{%
Samuel W. Akingbade\authororcid{0000-0002-6515-4073}%
}

\address{Department of Mathematics, University of Michigan, Ann Arbor, MI, USA.}
\email{sakingba@umich.edu}
\email{sakingba@mail.yu.edu (permanent)}

\begin{abstract}
Financial markets exhibit temporal organization that is not fully captured by volatility measures or linear correlation structure. We study a null-validated topological approach for quantifying financial market complexity using Bitcoin daily log returns and the S\&P 500 index as examples of cryptocurrency and broad U.S. equity market dynamics. The analysis uses the $L^1$ norm of the persistence landscapes computed from sliding-window delay embeddings. This quantity co-moves strongly with stochastic volatility during periods of market stress, but the strength and form of this relationship vary over time and differ between the two markets. Surrogate-based null models provide statistical validation of these observations. Rejection of shuffle surrogates rules out explanations based on marginal distributions alone, while departures from phase randomized surrogates indicate sensitivity to nonlinear and phase-dependent temporal organization beyond linear correlations. These results demonstrate that persistence landscape norms provide complementary information about market dynamics across market conditions.
\end{abstract}

\maketitle
\section{Introduction}
\label{sec:introduction}
Understanding the complex temporal dynamics of financial markets remains a central challenge in financial mathematics, quantitative finance, and econometrics. Traditional tools such as volatility estimates and linear correlation measures provide essential but limited perspectives on market behavior, particularly during periods of stress or structural change. Volatility, often modeled through stochastic processes, captures the scale of fluctuation but may fail to detect subtler dynamical features that reflect temporal organization or higher order dependencies in return series. As a result, there is growing interest in methods that can extract richer information about the geometry and temporal structure of financial time series beyond conventional statistical summaries.

In recent years, topological data analysis (TDA) has emerged as a promising framework for studying the shape of complex data. TDA seeks to describe intrinsic structural features that persist across multiple scales, leveraging tools from algebraic topology to quantify connectivity and loops in point clouds or time series embeddings. At the heart of TDA is persistent homology, a method for tracking the birth and death of topological features such as connected components and cycles as a scale parameter varies, yielding descriptors like barcodes and persistence diagrams that summarize data shape in a multiscale manner. Persistent homology has been applied across diverse domains where structure matters, including the analysis of complex biological and dynamical signals, due to its robustness to noise and coordinate-free characterization of data geometry; see, for example, \cite{Pereira2018Networks}.

In the context of time series, a common approach is to convert sequential observations into a delay or sliding window embedding that reconstructs the dynamics of the underlying process in a higher-dimensional space. This method draws on ideas related to Takens’ embedding theorem \cite{takens1981detecting} and has been studied in both theoretical and applied settings; by embedding a signal into a point cloud in a reconstructed state space, one can apply persistent homology to detect recurrent geometric patterns such as loops and voids that correspond to underlying dynamical phenomena \cite{perea2015sliding}. Practical summaries of these persistent features include persistence landscapes and their norms, which are scalar functions that quantify the prominence of topological features and enable statistical comparison across time.

Application of these methods to financial time series has grown in recent years. A major work by \cite{gidea2018topological} used persistent homology to analyze major U.S. stock indices, showing that persistence landscape norms increase before and during market meltdowns, thereby providing a novel signal that goes beyond standard volatility measures. More recently, \cite{AkingbadeGideaManziNateghi2024} offered a theoretical explanation for this behavior by linking the growth of persistence landscape norms to log-periodic power law singularity dynamics, which are commonly used to model speculative bubbles and critical transitions. They showed that when a financial time series follows such dynamics, its delay coordinate embedding naturally exhibits structured oscillatory geometry, leading to the emergence of persistent loop-like features in the reconstructed state space. In a related line of work, \cite{gidea2020topological} developed a topological framework for recognizing critical transitions in cryptocurrency markets, including Bitcoin and Ethereum, in the period leading up to the 2018 market crash. Additional studies have applied topological methods to financial time series in a variety of settings, including early warning detection, critical transitions, bubble identification, sparse portfolio and structural analysis of market dynamics; see, for example, \cite{ArvanitisDetsis2024,Ismail2021,gidea2017topological,
GoelPasrichaKanniainen2026,GoelFilipovicPasricha2025,rivera2019topological,9309855}.

Subsequent research has applied topological methods to change point detection in financial markets, demonstrating that persistent homology features can align with major economic events and volatility regimes across different stock markets \cite{YaoLiWuYangWang2025}. Persistence landscape norms have been shown to correlate with volatility and uncertainty in financial markets \cite{rudkin2021uncertainty,Souto2023}, however the dynamical origin of this signal and its dependence on temporal ordering and nonlinear structure remain largely unexplored. These contributions illustrate the potential of persistent homology to capture aspects of market dynamics that are not fully accessible via conventional statistical tools.

Despite these advances, several gaps remain in the rigorous validation and interpretation of topological summaries in financial settings. In particular, it is important to distinguish genuine temporal organization from artifacts of marginal distributions or linear second-order dependence. Null models and surrogate data provide a principled way to make this distinction by preserving selected statistical properties of a time series while disrupting others. In financial time series analysis, ~\cite{Fotiadis2023} developed a surrogate-based significance framework for detecting structural breaks associated with nonlinear causal interactions, using resampled multivariate time series consistent with a null model that preserves marginal and linear dependence properties. In the context of topological data analysis, ~\cite{BobrowskiSkraba2023} studied null distributions for persistent homology and showed that persistence diagrams arising from random point clouds can exhibit universal distributional behavior after appropriate normalization, providing a foundation for statistical hypothesis testing in TDA. Another use of null model comparisons in a non-financial setting can be found in \cite{stolz2017persistent}. These works highlight the importance of null-model comparisons for interpreting nonlinear and topological summaries.

Addressing this need is especially relevant in markets characterized by regime shifts, volatility clustering, and structural instability. Cryptocurrency markets provide one such setting and have been widely studied due to their pronounced volatility, behavioral effects, and rapid structural changes~\cite{Gaies2023,He2023,Wang2024}. Bitcoin, as the largest and most liquid cryptocurrency, provides a natural example for examining these issues, while the S\&P 500 index provides a broad U.S. equity-market comparison. Our work contributes to this literature by introducing a null-validated topological analysis of Bitcoin log returns and S\&P 500 index returns that explicitly compares persistence-landscape norms to surrogate ensembles designed to preserve specific statistical properties of the series while selectively disrupting temporal ordering or nonlinear and phase-dependent structure.

In this study, we investigate the geometric and temporal structure of financial return dynamics through a topological lens, treating the $L^1$ norm of the persistence landscapes computed from sliding window delay embeddings as a quantitative summary of reconstructed state-space geometry. The analysis is carried out for Bitcoin daily log returns and for the S\&P 500 index, providing examples of cryptocurrency and broad U.S. equity-market dynamics. Rather than interpreting this quantity solely as a proxy for market stress, we examine how it relates to and differs from established scale-based models of market variability. To this end, a central contribution of this work is a systematic comparison between the persistence landscape norm and filtered stochastic volatility estimates, including an analysis of their time varying association through rolling correlations. This comparison allows us to characterize regimes in which topological structure and volatility align, as well as regimes in which they decouple.

The second central contribution of this work is the incorporation of surrogate-based null models to assess the statistical origin of the observed topological signal. By employing both shuffle surrogates, which preserve the marginal distribution of returns while destroying temporal ordering, and phase-randomized surrogates, which preserve linear second order structure while removing nonlinear and phase-dependent dependencies, we disentangle the respective roles of marginal effects, linear correlations, and higher order temporal organization. This framework enables us to distinguish genuine geometric structure in the reconstructed dynamics from artifacts induced by distributional or linear properties of the time series.

Our results show that while the persistence landscape norm co-moves with stochastic volatility during periods of market stress, the dependence between the two is not stationary and differs across markets. In Bitcoin, the topology-volatility relationship is more unstable and exhibits a clearer regime shift, whereas in the S\&P 500 the coupling is more consistently positive but still time-varying. In both cases, surrogate null models show that the observed topological signal cannot be explained solely by marginal return distributions or by linear second order dependence. Together, these findings position the persistence landscape norm as a null-validated descriptor of financial time series that is sensitive to volatility while also capturing complementary information about temporal organization.

The structure of the paper is as follows. Section~\ref{sec:background} reviews the topological framework, including persistent homology, persistence landscapes, and their application to financial time series through sliding window and delay embeddings. Section~\ref{sec:sv_model} compares the $L^1$ norm of the persistence landscape for Bitcoin log returns with filtered stochastic volatility and examines their rolling correlation. Section~\ref{sec:surrogate} introduces surrogate-based null models for statistical validation of the persistence landscape norm as a descriptor of market dynamics beyond marginal distributional effects and linear second order dependence. Section~\ref{sec:sp500} applies the same framework to the S\&P 500 index as a broad U.S. equity market example and compares the resulting topology-volatility relationship and surrogate-null behavior with the Bitcoin application. Conclusions and future directions are presented in Section~\ref{sec:conclusions}.

\section{Background}
\label{sec:background}
\subsection{Persistent homology}
\label{sec:tda_method}

In this section, we describe the topological framework used throughout the paper, based on persistent homology. The tool provides a multiscale characterization of the geometry of point clouds derived from time series and allows us to extract robust numerical summaries of their evolving structure via persistence landscapes. The mathematical foundations of persistent homology and persistence landscapes are well established; see, for example, \cite{edelsbrunner2022computational,bubenik2015statistical,
bubenik2017persistence,Bubenik2018Persistence}.

\subsubsection{Persistent homology} It associates to a finite point cloud a family of topological spaces indexed by a resolution parameter, and tracks the evolution of topological features, such as connected components and loops, across scales. Persistence landscapes provide a functional representation of this multiscale information in a Banach space, enabling numerical summaries and statistical comparison. In this work, we ultimately summarize each point cloud by a single scalar quantity given by the $L^1$ norm of its persistence landscape.

Let
\[
X = \{x_0, x_1, \ldots, x_{m-1}\} \subset \mathbb{R}^N
\]
be a finite point cloud embedded in Euclidean space. To associate a topological space to $X$, we consider the Vietoris--Rips construction. For a fixed resolution parameter $t > 0$, the \emph{Vietoris--Rips simplicial complex} $VR(X,t)$ is defined by including a $k$-simplex with vertices $\{x_{i_0},\ldots,x_{i_k}\}$ whenever the pairwise distances between all vertices are strictly less than $t$, that is,
\[
d^X(x_{i_j}, x_{i_{j'}}) < t
\quad \text{for all } x_{i_j}, x_{i_{j'}} \in \{x_{i_0},\ldots,x_{i_k}\}.
\]
Intuitively, simplices are added whenever their vertices become indistinguishable at resolution $t$. As $t$ increases, these complexes form a nested sequence
\[
VR(X,t) \subseteq VR(X,t') \quad \text{for } t < t',
\]
which constitutes a \emph{filtration} of simplicial complexes.

At each resolution level, we compute the simplicial \emph{homology groups}
\[
H_n(VR(X,t))
\]
with coefficients in a fixed field, here taken to be $\mathbb{Z}_2$. The generators of these groups correspond to $n$-dimensional topological features: connected components for $n=0$, loops for $n=1$, voids for $n=2$, and so forth. In this work we restrict attention to one-dimensional homology, $n=1$, so that the analysis focuses exclusively on loop-like geometric structures in the embedded data.

A filtration of simplicial complexes induces a corresponding sequence of homology groups,
\[
H_n(VR(X,t)) \to H_n(VR(X,t')) \quad \text{for } t < t',
\]
via canonical homomorphisms induced by the inclusions in the filtration. The resulting collection is called a \emph{persistence module}. Persistence modules are uniquely decomposable into a direct sum of interval modules up to permutation \cite{ZomorodianCarlsson2005}. The multiset of intervals in this decomposition is called the barcode, and each interval represents the lifespan of a topological feature.

\subsubsection{Persistence diagrams} A homology class $\alpha \in H_n$ is said to be born at scale $b_\alpha$ if it first appears at $t:=b_\alpha$ and is not in the image of any class from smaller scales. The class persists across intermediate resolutions and is said to die at scale $d_\alpha > b_\alpha$ if its image becomes trivial at $t:=d_\alpha$. Each such class is therefore associated with a birth--death pair $(b_\alpha,d_\alpha)$, with finite multiplicity determined by the number of classes sharing the same birth and death values. The collection of all birth--death pairs arising from $H_n$ is encoded in the \emph{persistence diagram} $P_n$, which is a locally finite multiset of points supported on $U:=\lbrace (t_1,t_2)\in \mathbb{R}^2:t_1<t_2\rbrace$ together with points on the diagonal $\delta U:= \lbrace (t,t)\in \mathbb{R}^2\rbrace$ counted with infinite multiplicity, representing trivial features. The horizontal axis corresponds to birth values and the vertical axis to death values.

\subsubsection{Persistence landscapes} To obtain a functional representation suitable for numerical analysis, we map persistence diagrams into a Banach space using persistence landscapes. 

For each off-diagonal point $(b_\alpha,d_\alpha) \in P_n$, we define the piecewise linear function
\begin{equation}
f_{(b_\alpha,d_\alpha)}(x) =
\begin{cases}
x - b_\alpha, & x \in \left(b_\alpha, \tfrac{b_\alpha+d_\alpha}{2}\right], \\
d_\alpha - x, & x \in \left(\tfrac{b_\alpha+d_\alpha}{2}, d_\alpha\right), \\
0, & \text{otherwise}.
\end{cases}
\end{equation}
Given a persistence diagram with finitely many off-diagonal points, the \emph{persistence landscape} is defined as the sequence of functions
\begin{equation}
\lambda_n = (\lambda_n(i))_{i \in \mathbb{N}},
\quad
\lambda_n(i)(x) = i_{\max}\{ f_{(b_\alpha,d_\alpha)}(x) \mid (b_\alpha,d_\alpha) \in P_n \},
\end{equation}
where $i_{\max}$ denotes the $i$-th largest value among the indicated functions. If the $i$-th largest value does not exist, $\lambda_n(i)(x)$ is set to zero. The resulting object lies in the Banach space $L^p(\mathbb{N} \times \mathbb{R})$ for any $p \geq 1$.

For a persistence landscape $\lambda_n$, the \emph{$L^p$ norm} is defined by
\begin{equation}
\|\lambda_n\|_p =
\left( \sum_{i=1}^\infty \|\lambda_n(i)\|_p^p \right)^{1/p},
\quad
\|\lambda_n(i)\|_p = \left( \int_{\mathbb{R}} |\lambda_n(i)(x)|^p \, dx \right)^{1/p}.
\end{equation}
In this work, we focus exclusively on the $L^1$ norm. For finite persistence diagrams, this norm admits the closed-form expression \cite{bubenik2015statistical}
\begin{equation}
\|\lambda_n\|_1 = \frac{1}{4} \sum_{\alpha} (d_\alpha - b_\alpha)^2.
\end{equation}
In practice, we compute a numerically approximated and truncated version of this quantity by integrating the first $i_{\max}$ landscape layers over a finite grid of scale values.

\subsection{Application to time series}
\label{sec:TDA_time_series}
We now describe how this framework is applied to financial time series following \cite{AkingbadeGideaManziNateghi2024,
gidea2020topological,ThesisAkingbade2024}.

Let
\[
Z = \{z_0, z_1, \ldots, z_{N-1}\}
\]
be a real valued time series of length $N$. Fix an embedding dimension $m$ and a time delay $d \geq 1$. The delay-coordinate embedding produces a sequence of vectors
\[
x_t = (z_t, z_{t+d}, \ldots, z_{t+(m-1)d}),
\]
defined for all indices $t$ such that the coordinates exist. 

\subsubsection{Choice of embedding dimension and delay.}

The use of delay-coordinate embeddings in this work is motivated by classical results from nonlinear dynamical systems theory, which establish conditions under which the geometry of an underlying state space can be reconstructed from time-delayed observations.

A foundational result is \emph{Takens’} embedding theorem \cite{takens1981detecting}. Consider a discrete-time dynamical system $x_{t+1} = T(x_t)$ evolving on a compact smooth manifold $M$ of dimension $D$, together with a smooth observation function $h : M \to \mathbb{R}$. Let $\{z_t\}$ be the scalar time series defined by $z_t = h(x_t)$. For a fixed delay $d \geq 1$, define the delay-coordinate map
\[
\Phi : M \to \mathbb{R}^m, \qquad
\Phi(x_t) = (z_t, z_{t+d}, \ldots, z_{t+(m-1)d}).
\]
Takens’ theorem states that, provided $m \geq 2D+1$, it is a generic property of the pair $(T,h)$ that $\Phi$ is an embedding, i.e., a smooth injective map with a smooth inverse onto its image. In this sense, the delay-coordinate vectors provide a faithful geometric representation of the underlying state space.

This result was subsequently generalized by \cite{tim1991embedology}, who showed that if the dynamics admit a compact attractor $A \subset M$ of finite fractal dimension $D$, then an embedding of the attractor can be achieved for $m \geq 2D+1$ under the weaker assumption of prevalence rather than genericity. This extension is particularly relevant in practical settings where the dynamics evolve on a lower dimensional invariant set.

In the present study, delay coordinate embeddings are employed not for exact state-space reconstruction, but as a geometric representation from which topological features can be extracted robustly via persistent homology. Consequently, the embedding dimension $m$ and delay $\tau$ are chosen empirically to be sufficient for capturing loop-like geometric structure in the delay embedding while remaining computationally tractable.

\subsubsection{Topological summaries of Bitcoin log return dynamics} To detect temporal variation in topological signal, we apply a sliding window of length $w$ to the sequence $\{x_t\}$, yielding a time-indexed family of point clouds
\[
X^t = \{x_t, x_{t+1}, \ldots, x_{t+w-1}\}.
\]

For each window $X^t$, we compute the Vietoris--Rips filtration, extract the one dimensional persistence diagram, construct the corresponding persistence landscape, and evaluate the truncated $L^1$ landscape norm.

\begin{figure}[htbp]
\includegraphics[width=.9\textwidth]{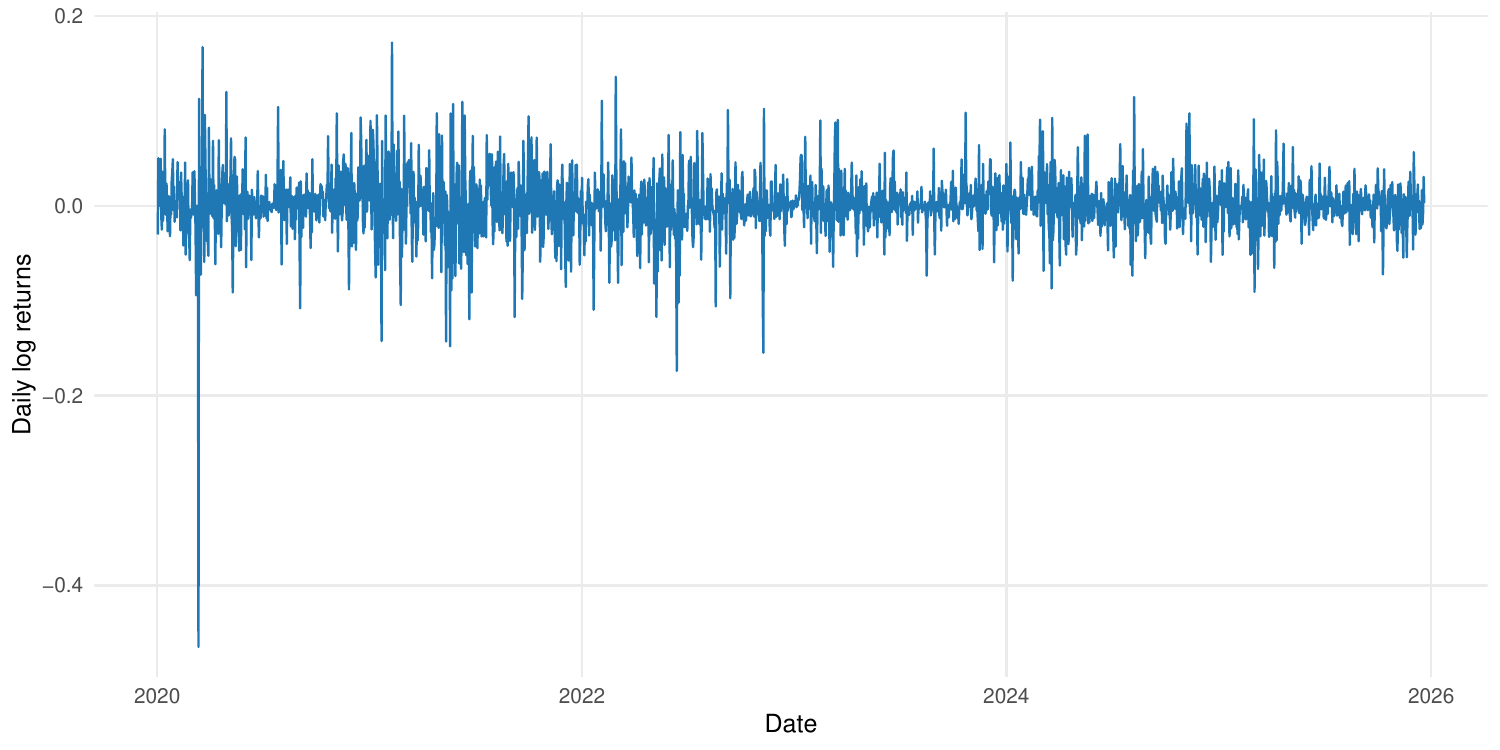}
\caption{Bitcoin daily log returns from January 1, 2020 to December 20, 2025.}
\label{fig:logreturns}
\end{figure}

Bitcoin price data from January 1, 2020 to December 20, 2025 were obtained from Yahoo Finance \cite{Yahoofinance} and used to compute daily log returns shown in Figure \ref{fig:logreturns}. The topological data analysis procedure described above is then applied to the standardized Bitcoin log returns.

The result is a scalar time series
\[
t \mapsto \|\lambda^t\|_1,
\]
which quantifies the evolving prominence of loop-like geometric structure in the embedded dynamics.

\begin{figure}[htbp]
\includegraphics[width=.9\textwidth]{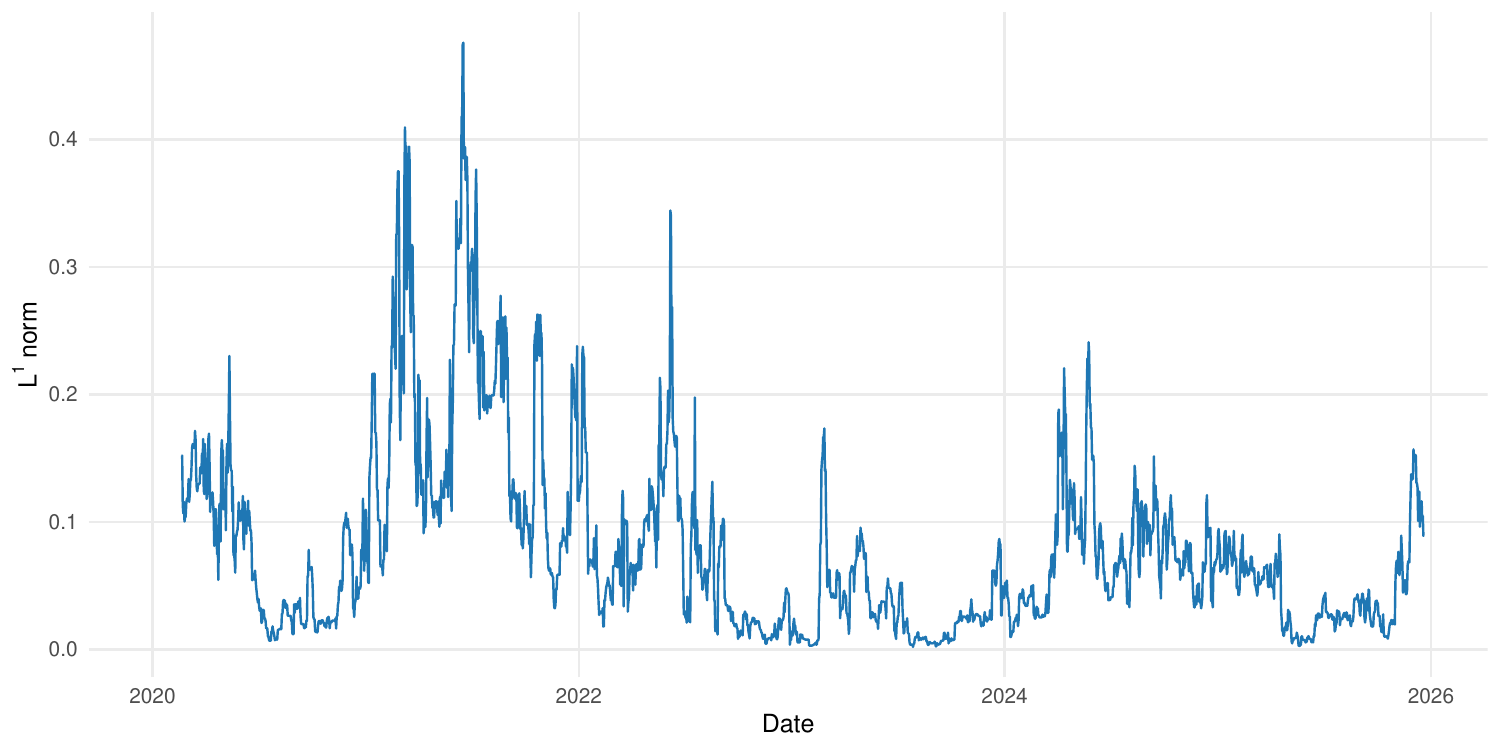}
\caption{$L^1$ norm of the persistence landscape of Bitcoin log returns for $w=50, m=4, \tau=2$.}
\label{fig:L1_of_returns}
\end{figure}

We show the time evolution of the $L^1$ norm of the persistence landscape computed from sliding window delay embeddings of standardized Bitcoin log returns in Figure~\ref{fig:L1_of_returns}, using an embedding dimension of $4$, a delay of $2$ and a sliding window of length $50$.

The resulting series exhibits multiple intervals of elevated $L^1$ values throughout the sample, reflecting the recurrence of loop-like geometric structure in the delay space across several distinct periods. These elevated intervals are interspersed with lower activity regimes, indicating that the strength of the topological signal varies over time rather than remaining uniformly high. This temporal variability motivates examining whether the observed topological signal can be explained by conventional market descriptors such as sentiment or volatility.

\section{Bitcoin stochastic volatility comparison}
\label{sec:sv_model}
\subsection{Stochastic volatility model}
We model the Bitcoin log return series using a standard stochastic volatility (SV) specification, in which the latent state corresponds to the log variance of returns. Let $h_t$ denote the latent log variance at time $t$. The model is defined by the following two equations.
\subsubsection{State (process) equation}
The latent log variance follows a mean reverting AR(1) process,
\begin{equation} h_t=\mu+\phi(h_{t-1}-\mu)+\sigma_{\eta}\eta_t, \ \ \ \ \ \eta_t \stackrel{\text{iid}}{\sim} \mathcal{N}(0,1). \end{equation}
where $\mu$ is the long run mean of the log variance, $\phi$ is the autoregressive coefficient, and $\sigma_\eta>0$ governs the magnitude of stochastic shocks to the latent log variance.

We restrict \(\phi \in (0,1)\) so that the latent log-variance process has a well-defined stationary distribution. This assumption is not intended to assert that the observed financial return series is globally stationary over the full sample, especially during periods of market instability. Rather, the stationary stochastic volatility model is used as a parsimonious latent-volatility benchmark against which the topological signal can be compared. Temporary volatility bursts and crash-related episodes are accommodated through innovations $\eta_t$ in the latent state and through the exponential mapping from log variance to conditional volatility.

\subsubsection{Observation (measurement) equation}
Conditional on $h_t$, returns are Gaussian with zero mean and variance $\exp(h_t)$,
\[
z_t \mid h_t \sim \mathcal{N}(0,\exp(h_t))
\]

Under this formulation, $\exp(h_t)$ represents the conditional variance of returns and $\exp(h_t/2)$ the conditional volatility. This specification corresponds to the canonical discrete-time stochastic volatility model widely used in empirical finance (e.g.,\cite{Taylor1986,Harvey1994,Kim1998,Shephard1996,HullWhite1987}).

\subsection{Inference and parameter estimation}
The SV model is cast as a partially observed Markov process (state-space model) and fitted using likelihood methods based on Monte Carlo sampling. Inference is carried out within the pomp framework, which supports likelihood evaluation and parameter estimation via simulation for nonlinear state-space models with non Gaussian noise.

To enforce parameter constraints during optimization, the model is reparameterized internally so that $\sigma_\eta$ is estimated on the log scale and $\phi$ on the logit scale, ensuring $\sigma_\eta>0$ and $\phi\in(0,1)$ throughout estimation.

Initial parameter values are chosen heuristically. In particular, the long run mean $\mu$ is initialized using the logarithm of the empirical variance of the return series, while the remaining parameters are set to plausible values reflecting high persistence in financial volatility.

Model parameters $\theta=(\mu,\phi,\sigma_\eta,h_0)$ are estimated by maximum likelihood using iterated filtering (IF2), a simulation-based algorithm for partially observed Markov processes \cite{Ionides2011,King2016}. Iterated filtering repeatedly applies a particle filter while perturbing parameters via a random walk with gradually decreasing variance, allowing the algorithm to ascend the likelihood surface of the state-space model.

For robustness against local maxima and Monte Carlo variability, the iterated filtering procedure was repeated 10 times from the same initial parameter guess. For each replicate, the likelihood at the final parameter estimate was evaluated 5 times using independent particle filters, and the replicate likelihood was summarized using a log-mean-exp aggregation. The parameter set associated with the highest aggregated likelihood was selected as the final estimate $\hat{\theta}$.

\subsection{Filtered latent volatility}
Given the estimated parameters \(\hat{\theta}\), we apply a particle filter to
approximate the sequence of filtering distributions
\[
p(h_t \mid z_{1:t}; \hat{\theta}), \qquad t = 1,\dots,n.
\]
The primary quantity of interest is the filtered posterior mean of the latent log variance,
\[
\hat{h}_t := \mathbb{E}[h_t \mid z_{1:t}; \hat{\theta}].
\]
In practice, this expectation is approximated using the particle representation
returned by the filter,
\[
\hat{h}_t \approx \frac{1}{N_p} \sum_{i=1}^{N_p} h_t^{(i)},
\]
where \(h_t^{(i)}\) denotes the \(i\)-th particle at time \(t\) after resampling.

\begin{figure}
\includegraphics[width=.9\textwidth]{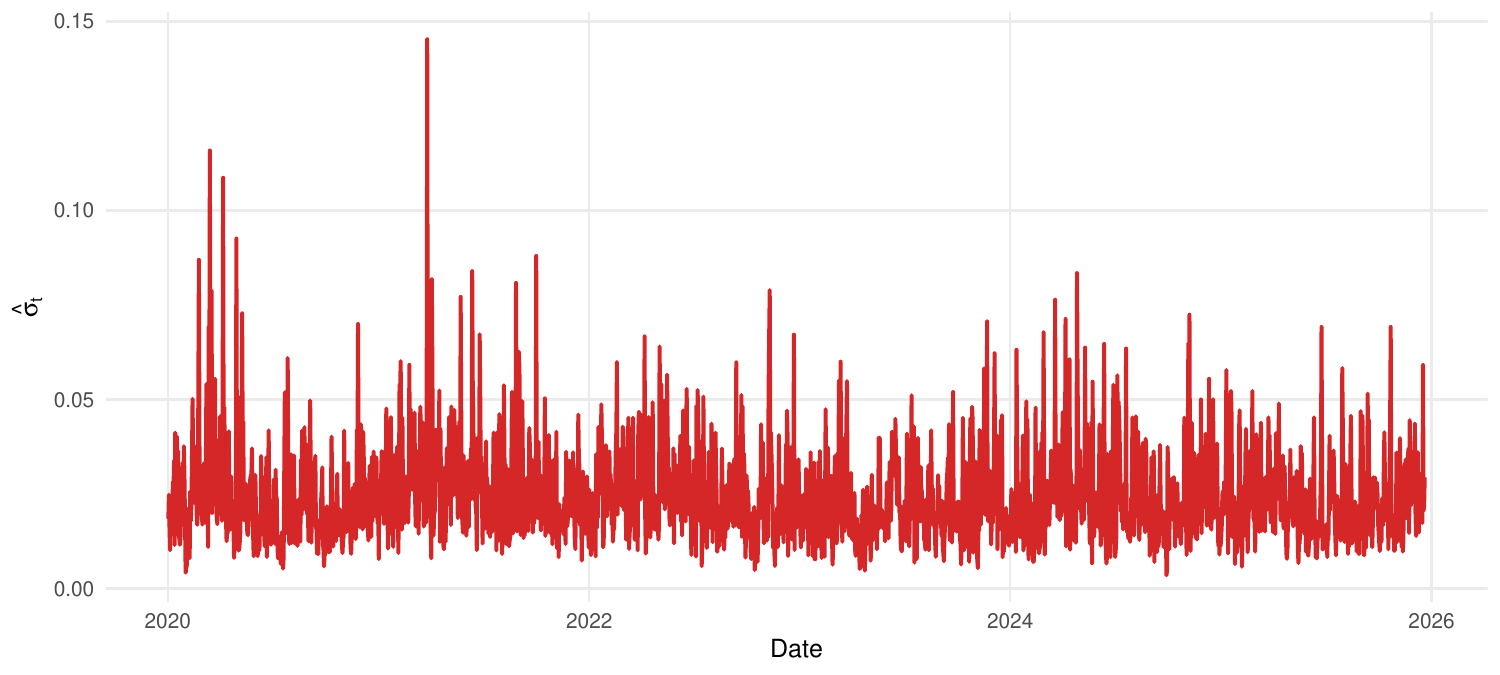}
\caption{Filtered conditional volatility estimate $\hat{\sigma}_t$ obtained from the stochastic volatility model of Bitcoin log returns.}
\label{fig:sv_sigma}
\end{figure}

From the filtered log variance sequence \(\{\hat{h}_t\}\), we construct estimates
of the conditional variance and volatility as
\begin{equation}
\hat{V}_t = \exp(\hat{h}_t),
\qquad
\hat{\sigma}_t = \sqrt{\hat{V}_t} = \exp\!\left(\frac{\hat{h}_t}{2}\right).
\end{equation}
The resulting series \((\hat{h}_t, \hat{V}_t, \hat{\sigma}_t)\) provides a
time-resolved estimate of latent stochastic volatility.
Figure~\ref{fig:sv_sigma} displays the filtered volatility series over calendar time, which serves as a latent volatility benchmark for subsequent analysis.

Having obtained a filtered estimate of latent stochastic volatility, we now empirically compare its temporal evolution with the $L^1$ norm of the persistence landscape computed from sliding-window embeddings of Bitcoin log returns.

\subsection{Empirical comparison with topological signal}

\begin{figure}[htbp]
\includegraphics[width=.9\textwidth]{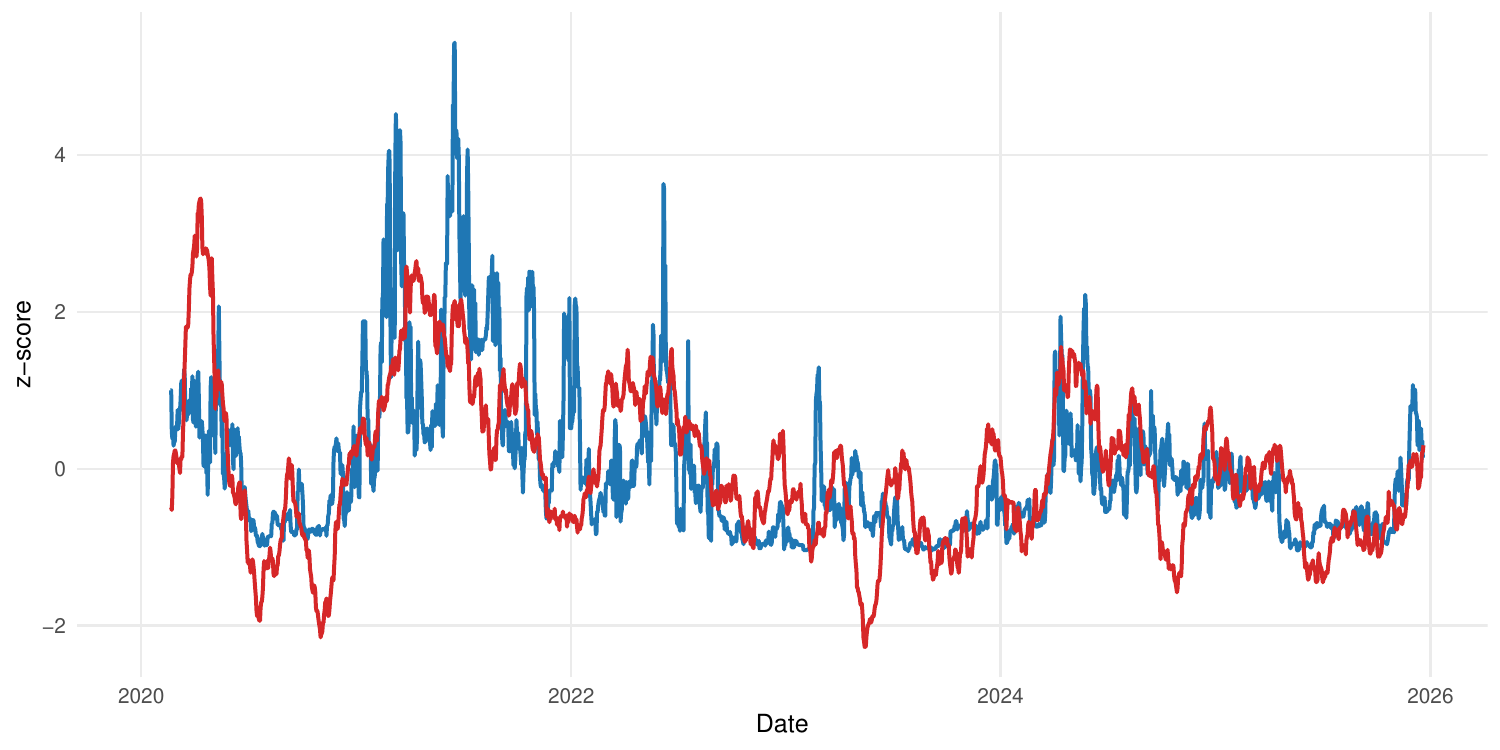}
\caption{Standardized comparison of the $L^1$ norm of the persistence landscape (blue) and filtered stochastic volatility (red).}
\label{fig:overlay_L1ret_vs_SV}
\end{figure}

In Figure \ref{fig:overlay_L1ret_vs_SV}, we present a standardized comparison between the $L^1$ norm of the persistence landscape computed from the Bitcoin log return series $\lbrace z_t\rbrace$ and the filtered stochastic volatility estimate. 

In early $2020$, the onset of the COVID-19 crisis is marked by an abrupt and extreme surge in stochastic volatility. During this period, the topological signal rises as well, indicating that the large amplitude fluctuations associated with the market crash are accompanied by pronounced geometric organization in the reconstructed return dynamics. This co-movement persists throughout much of $2020–2021$, a period characterized by repeated volatility spikes and sustained market stress, during which both measures remain elevated and closely aligned.

Following these stress episodes, the series enter multiple phases of volatility compression. In mid $2020$ and again in late $2020$, the standardized volatility drops to values near $-2$, indicating unusually low volatility relative to its long-run mean. During these intervals, the topological signal also declines below its mean, but typically not to the same extent as volatility. Furthermore, following the late $2021$ to early $2022$ decline in market volatility, the stochastic volatility estimate remains suppressed, whereas the topological signal continues to display intermittent bursts and sustained deviations from baseline. This asymmetric relaxation suggests that while fluctuation amplitude contracts rapidly after major shocks, geometric organization in return dynamics decays more gradually, retaining some structure even as volatility becomes unusually subdued.

A similar but more prolonged compression regime occurs in $2023$. In standardized units, stochastic volatility remains persistently suppressed for an extended period, reaching values near $-2$ and exhibiting relatively little variation. Over the same interval, the topological signal is often negative as well, but again does not compress as strongly and continues to display intermittent excursions. This behavior indicates that periods of apparent market calm, as measured by volatility, can still be associated with nontrivial temporal organization in returns, though at a reduced level compared to crisis periods.

All these observations indicate two key features. First, during periods of severe market stress, volatility and topological signal rise together reflecting the joint presence of large fluctuations and strong geometric structure. Second, outside these regimes, particularly during volatility compression phases, the relationship becomes asymmetric: volatility often contracts more strongly and more rapidly than the topological signal. This indicates that the $L^1$ norm is sensitive not only to fluctuation scale but also to the persistence and organization of return dynamics, which can survive, albeit in weakened form, during extended periods of low volatility.

\begin{figure}
\includegraphics[width=.9\textwidth]{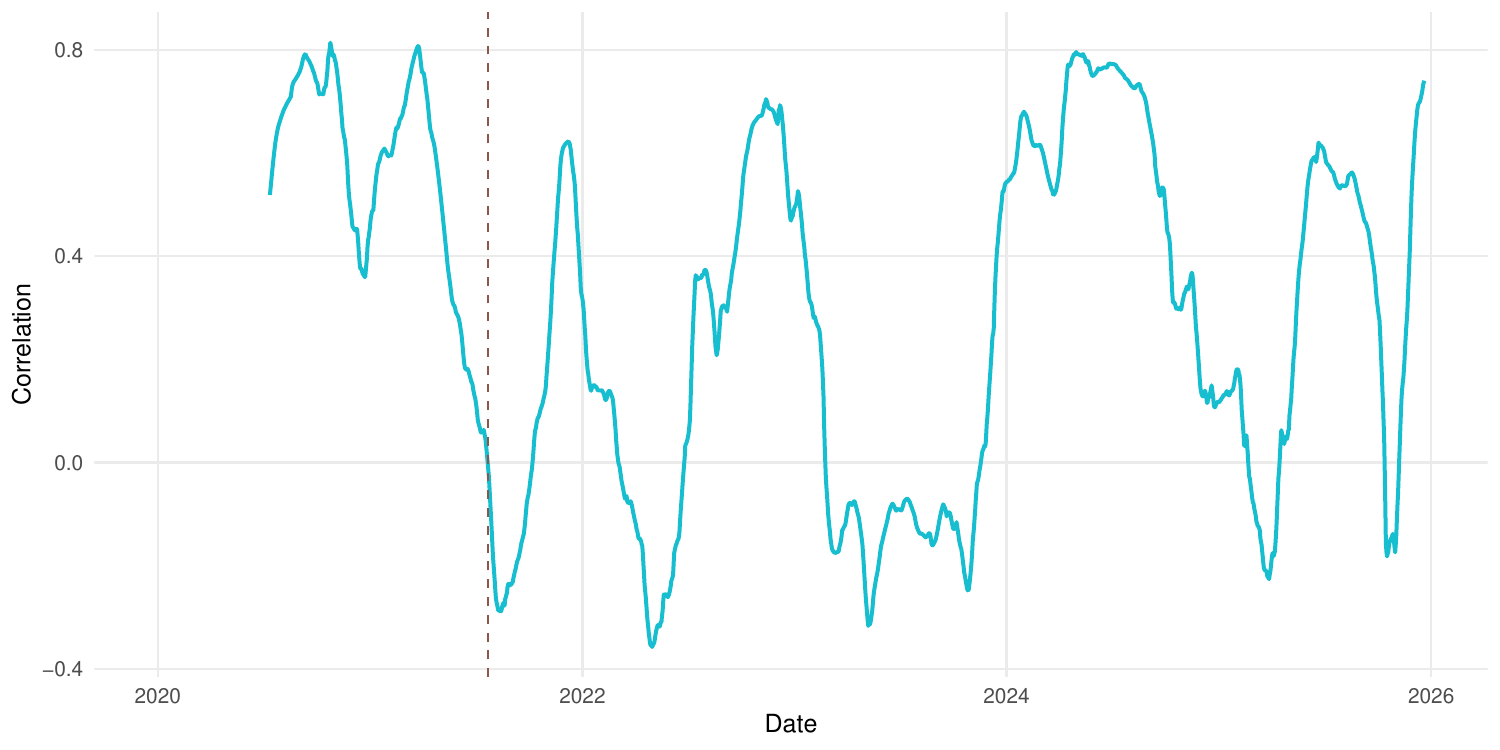}
\caption{Rolling correlation between the $L^1$ norm and stochastic volatility (180-day window).}
\label{fig:rollcorr_L1_SV}
\end{figure}

To quantify the evolving relationship between topological signal and volatility, we examine the rolling correlation between the $L^1$ norm and the filtered stochastic volatility estimate over a 180-day horizon in Figure~\ref{fig:rollcorr_L1_SV}. At each time point, the curve reports the linear association between topology and volatility over the preceding six months, providing a time-resolved measure of their dependence.

During periods of heightened market stress, the rolling correlation attains sustained positive values, indicating that increases in volatility are accompanied by concurrent increases in topological signal. This behavior is consistent with the co-movement observed in Figure~\ref{fig:overlay_L1ret_vs_SV} during crisis regimes, where large-amplitude fluctuations and strong geometric organization arise together.

In contrast, following the subsequent contraction in volatility, the rolling correlation undergoes a pronounced decline and becomes markedly unstable, with extended intervals of weak or even negative association. Although the correlation exhibits multiple local minima, changepoint detection using the Pruned Exact Linear Time (PELT) algorithm identifies a single dominant shift (indicated by the brown dashed line in Figure~\ref{fig:rollcorr_L1_SV}) in the mean level of dependence. This transition separates an earlier regime characterized by persistently strong coupling from a later regime in which the association between topology and volatility is weaker and more variable.

This behavior indicates that the dependence between topological signal and volatility is not stationary over time. While the two measures align closely during crisis periods, the topological signal may persist or exhibit intermittent intensification even as volatility subsides. This further supports the interpretation that the $L^1$ norm captures aspects of return dynamics beyond fluctuation scale alone, reflecting changes in temporal organization that are not fully encoded by volatility.

\subsection{Topological signal beyond volatility and sentiment}

\begin{figure}
\includegraphics[width=.9\textwidth]{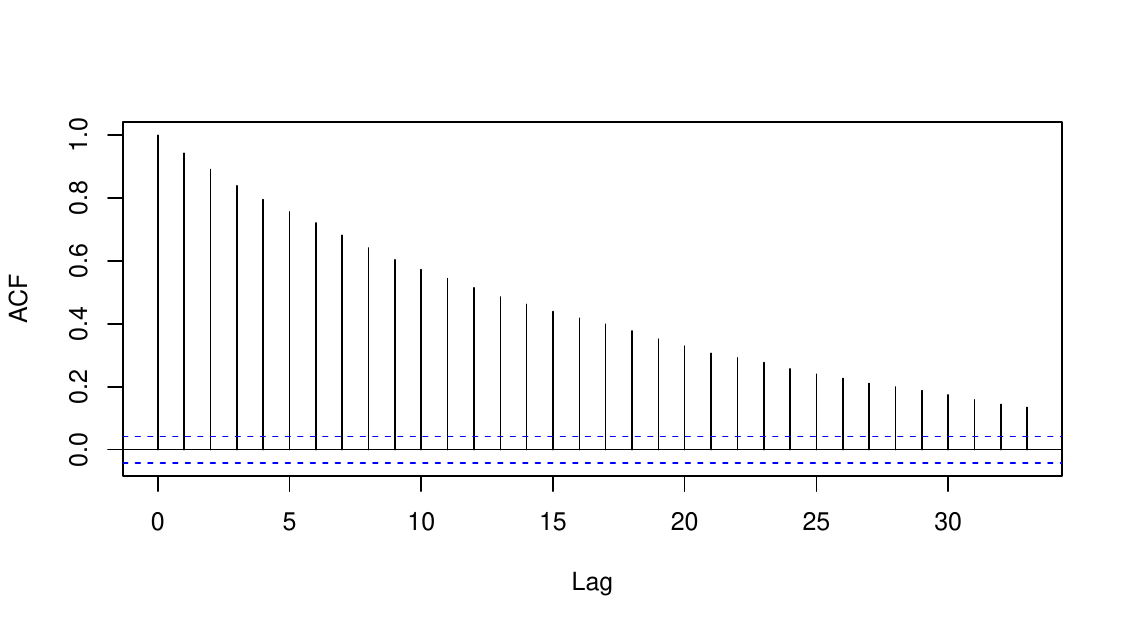}
\caption{ACF of the residual $L^1$ norm of the persistence landscape computed from Bitcoin log return embeddings after removing linear effects of stochastic volatility and sentiment.}
\label{fig:acf_resid_L1_ret}
\end{figure}

We perform a residualization analysis to examine whether the Bitcoin topological signal is fully explained by stochastic volatility and coarse sentiment variation. Sentiment is measured using the Crypto Fear \& Greed Index obtained via the public API provided by alternative.me \cite{cryptoalternative}. This index is a composite sentiment indicator scaled between 0 (Extreme Fear) and 100 (Extreme Greed), designed to capture aggregate market sentiment by combining multiple behavioral and market-based inputs, including price momentum, volatility, trading volume, and social media activity. 

In this analysis, the $L^1$ norm of the persistence landscape computed from delay embeddings of Bitcoin log returns is regressed on contemporaneous measures of volatility and sentiment.
Specifically, we consider the linear model

\begin{equation}
L_1(t)
=
\beta_0
+
\beta_1 \, \overline{\sigma}_t
+
\beta_2 \, \overline{F}_t
+
\beta_3 \, S_t
+
\varepsilon_t ,
\end{equation}

where $L_1(t)$ denotes the $L^1$ norm summarizing the topological signal of the log return dynamics, $\overline{\sigma}_t$ is the rolling window mean of the filtered stochastic volatility, $\overline{F}_t$ is the rolling window mean of the Fear \& Greed sentiment index, and $S_t:=\sqrt{Var(F_{t-L+1:t})}$ denotes the rolling window standard deviation of sentiment. The coefficients $\beta_0,...,\beta_3$ capture linear associations between the topological summary and the covariates, while the residual term $\varepsilon_t$ represents the component of topological variation not explained by volatility or sentiment.

The primary object of interest is the temporal dependence structure of the residual series $\varepsilon_t$. Figure~\ref{fig:acf_resid_L1_ret} displays the autocorrelation function (ACF) of $\varepsilon_t$, which exhibits a slow decay with statistically significant correlations persisting across many lags. This behavior is inconsistent with a white noise residual and indicates that structured temporal organization remains in the topological signal after removing linear effects of volatility and coarse sentiment statistics.

\section{Bitcoin surrogate null models}
\label{sec:surrogate}
Let $\lbrace z_t\rbrace_{t=1}^N$ denote the standardized Bitcoin log return series. To determine whether the $L^1$ norm of the persistence landscape obtained from sliding window delay embeddings of $\lbrace z_t\rbrace$ arise from genuine temporal structure, rather than effects attributable to the marginal distribution or linear second-order dependence, we compare the resulting time series of persistence landscape norms to those obtained from two classes of surrogate Bitcoin log return processes. Each surrogate is transformed through the same embedding and persistent homology pipeline, thereby inducing a null distribution for the windowed persistence landscape norm defined in \ref{sec:tda_method}.

\subsection{Shuffle (i.i.d.) surrogate}\label{subsec:shuffle-surrogate}

Let $\pi$ be a random permutation of $\{1,\dots,N\}$ drawn uniformly from the symmetric group $S_N$.
The \emph{shuffle surrogate} of the observed series $\{z_t\}_{t=1}^N$ is defined by
\begin{equation}
z_t^{s} := z_{\pi(t)}, \qquad t=1,\dots,N .
\end{equation}
This construction preserves the empirical marginal distribution of $\{z_t\}$ exactly, but destroys all temporal ordering and hence all serial dependence. In probabilistic terms, this surrogate corresponds to a null in which the observed series is treated as exchangeable.

For each realization $j=1,\dots,N_s$, we generate an independent permutation $\pi_j$, compute the corresponding surrogate series $z_t^{s,j}$,
and apply Section \ref{sec:tda_method} to obtain a surrogate topological time series.

\subsection{FFT phase-randomized surrogate}\label{subsec:fft-surrogate}
The second null preserves second order temporal structure while destroying nonlinear and phase dependent features.
Let $\tilde z_t = z_t - \bar z$ denote the demeaned series, and define its discrete Fourier transform (DFT) by
\begin{equation}
Z_k := \sum_{t=0}^{N-1} \tilde z_t \, e^{-2\pi i k t/N}, \qquad k=0,\dots,N-1 .
\end{equation}
Write $Z_k = |Z_k|e^{i\theta_k}$. A \emph{phase randomized} surrogate is constructed by replacing the phases $\theta_k$ at positive frequencies with i.i.d.\ random variables
\begin{equation}
\phi_k \sim \mathrm{Unif}(0,2\pi),
\end{equation}
while retaining the magnitudes $|Z_k|$, and enforcing conjugate symmetry so that the inverse transform is real-valued. Concretely, for $k=1,\dots,\lfloor (N-1)/2\rfloor$, define
\begin{equation}
Z_k^{\mathrm{fft}} := |Z_k|e^{i\phi_k}, 
\qquad 
Z_{N-k}^{\mathrm{fft}} := \overline{Z_k^{\mathrm{fft}}},
\end{equation}
and set $Z_0^{\mathrm{fft}} := Z_0$. If $N$ is even, the Nyquist component $k=N/2$ must be real; one may set $Z_{N/2}^{\mathrm{fft}} := Z_{N/2}$ (or equivalently randomize with a sign change consistent with real-valuedness).

The inverse DFT yields a surrogate demeaned series
\begin{equation}
\tilde z_t^{\mathrm{fft}} := \frac{1}{N}\sum_{k=0}^{N-1} Z_k^{\mathrm{fft}} e^{2\pi i k t/N}, \qquad t=0,\dots,N-1,
\end{equation}
to which the original mean $\bar z$ is restored via $z_t^{\mathrm{fft}} = \tilde z_t^{\mathrm{fft}} + \bar z$.

This procedure preserves the power spectrum (and hence the autocovariance function) asymptotically, while destroying higher-order temporal dependencies and nonlinear phase relationships \cite{SchreiberSchmitz2000,theiler1992testing}. As before, for each realization $j=1,\dots,N_{\mathrm{fft}}$, we apply \ref{sec:tda_method} to obtain a surrogate topological time series.

\subsection{Pointwise null envelopes for persistence landscape norms}\label{sec:null-envelopes}
For each sliding window and each null model, we generate a surrogate ensemble of persistence landscape norms by applying the same embedding and topological pipeline described in Section~\ref{sec:tda_method} to surrogate versions of $\{z_t\}$. These surrogate values form an empirical sample from the null distribution of the window-level statistic. In all experiments, null envelopes are estimated using 30 surrogate realizations per null model. The qualitative behavior of the envelopes and exceedance patterns is robust to further increases in the number of surrogate realizations.

At each window, we summarize this null distribution by computing its mean, as well as its 5th and 95th empirical percentiles. The interval defined by these two percentiles constitutes a pointwise null envelope for the persistence landscape norm at that window.

This envelope provides a marginal comparison at each window index and should not be interpreted as a simultaneous confidence band across time; in particular, it does not control the family-wise error rate over all windows.

\subsection{Exceedance statistics}\label{sec:exceedance}
We investigate deviations of the persistence landscape norm from its surrogate-based null envelopes by examining exceedances across sliding windows. We restrict attention to windows for which both the norm computed from the Bitcoin log return series and the corresponding null quantiles are well defined.

An exceedance below the null is recorded when the persistence landscape norm falls below the lower (5th percentile) bound of the null envelope, while an exceedance above the null is recorded when it exceeds the upper (95th percentile) bound. We count the total number of such lower and upper exceedances across all valid windows.

Finally, we report the fractions of windows in which the persistence landscape norm lies below or above the null envelope, respectively. These fractions provide a concise summary of how often the observed topological signal deviates from that expected under the surrogate null models.

The empirical behavior of these envelopes and the associated exceedance frequencies are examined in Section \ref{sec:null-interpretation} and summarized in Table~\ref{tab:null_exceedance}.

\subsection{Comparison with surrogate-based null envelopes}\label{sec:null-interpretation}
\subsubsection{Shuffle surrogate comparison}

\begin{figure}
\includegraphics[width=.9\textwidth]{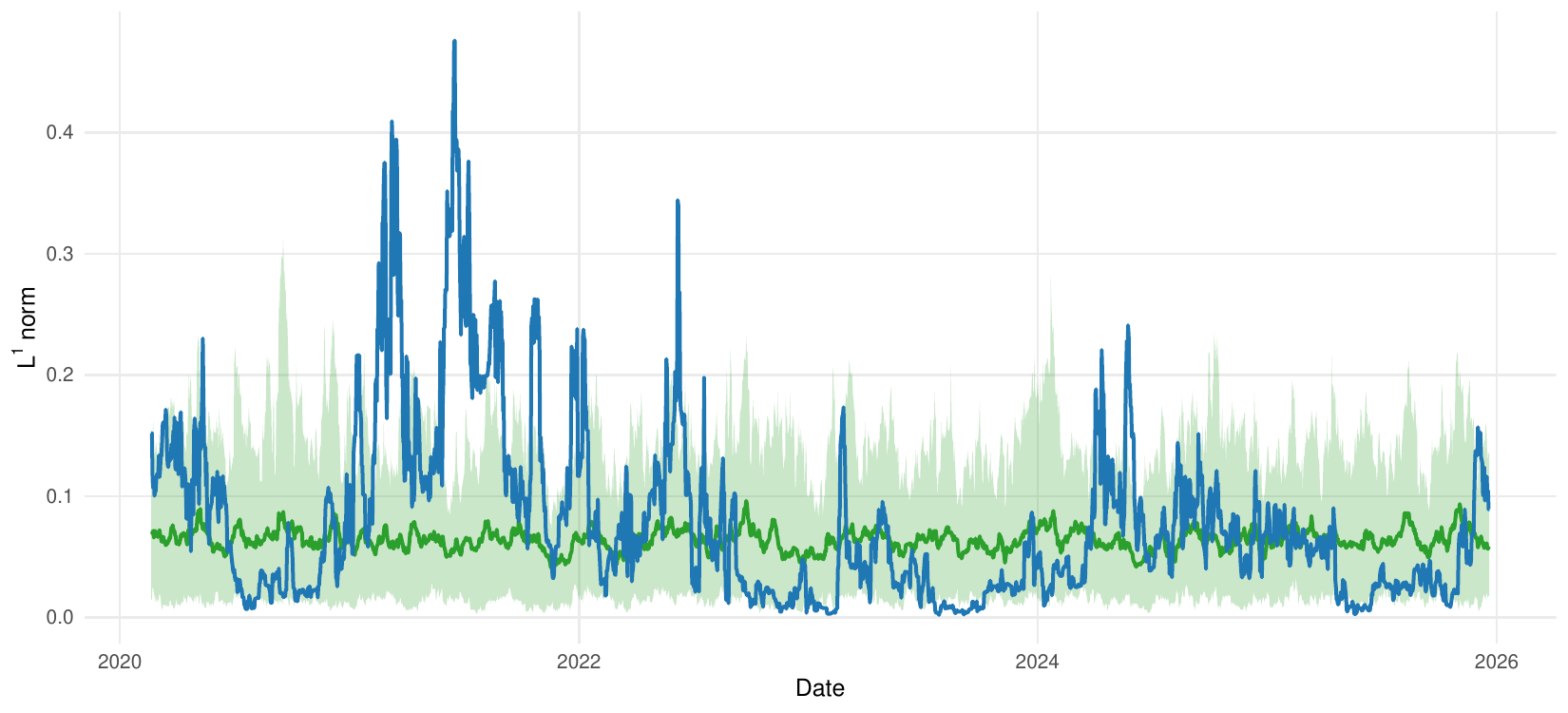}
\caption{Comparison of the observed $L^1$ norm of the persistence landscape for Bitcoin (blue) with pointwise null envelopes obtained from shuffle surrogates.}
\label{fig:shuffle}
\end{figure}

In Figure \ref{fig:shuffle}, we compare the observed time series of $L^1$ norm of the persistence landscape to pointwise null envelopes obtained from shuffle surrogates. Because shuffle surrogates preserve the marginal distribution of returns but destroy all temporal ordering, exceedances relative to this envelope indicate sensitivity of the windowed norms to temporal dependence.

The series frequently exceeds the upper bound of the shuffle null envelope, particularly during pronounced volatility episodes in early $2021$ and mid $2022$, where the landscape norm rises well above what is expected under random reordering of returns. These excursions indicate that the topological signal captured by the sliding-window embeddings is not explained by the empirical distribution of returns alone and depends critically on temporal ordering.

At the same time, there are periods, most notably during relatively calm market regimes, where the norm of the persistence landscape falls within or below the shuffle envelope, suggesting that during these intervals the windowed topological signal is closer to that expected under weak or absent temporal dependence.

The exceedance rates confirm that a non-negligible fraction of windows lie above the 95\% shuffle envelope, with rejection frequencies well above the nominal 5\% level. This supports rejection of the hypothesis that the norm of the persistence landscapes arises solely from the marginal distribution of Bitcoin log returns.

\subsubsection{FFT phase-randomized surrogate comparison}
\begin{figure}
\includegraphics[width=0.9\textwidth]{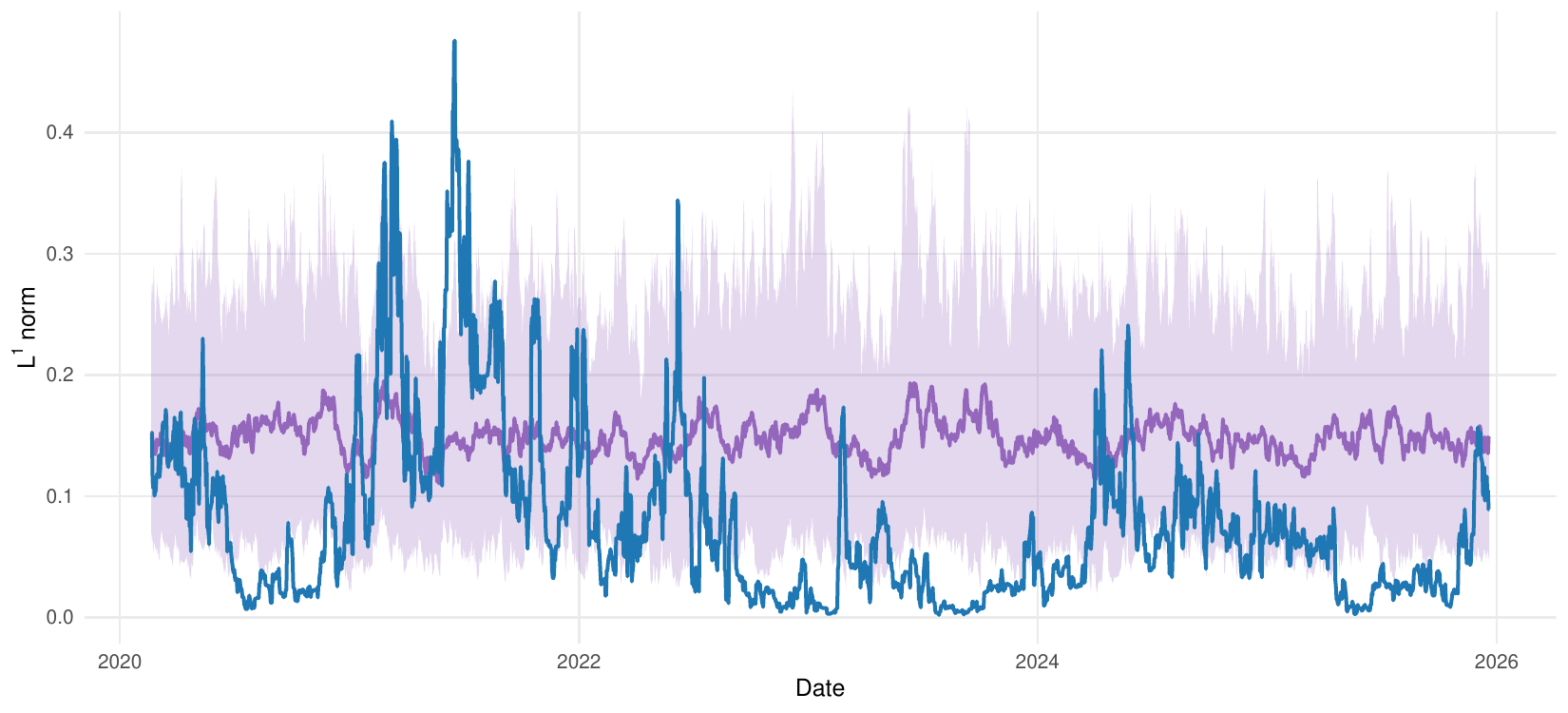}
\caption{Comparison of the observed $L^1$ norm of the persistence landscape for Bitcoin (blue) with pointwise null envelopes generated by FFT phase-randomized surrogates.}
\label{fig:FFT}
\end{figure}

In Figure \ref{fig:FFT}, we compare the observed persistence landscape norm to null envelopes generated by FFT phase randomized surrogates. These surrogates preserve the power spectrum and hence linear second order temporal structure while destroying nonlinear and phase-dependent dependencies.

Relative to this more restrictive null, the norm exhibits a different pattern. While occasional excursions above the FFT envelope occur, they are markedly fewer and more localized than in the shuffle case. In many periods, particularly after 2022, the series lies persistently below the mean of the FFT null ensemble, and often near the lower edge of the envelope.

This behavior indicates that the norm of the persistent landscape is sensitive to structure beyond linear correlation alone. The suppression of the norm relative to the FFT null suggests that nonlinear temporal organization present in $\lbrace z_t\rbrace$ constrains the geometry of the delay embedded point clouds in a way that is not reproduced by phase randomized surrogates.

Both of these results demonstrate that the $L^1$ norm of the persistence landscape derived from Bitcoin log returns exhibits temporal structure that cannot be explained by either the marginal distribution alone or by linear second-order temporal dependence. The shuffle surrogate analysis establishes sensitivity to temporal ordering, while the FFT phase randomized analysis indicates structure beyond that captured by linear correlations.

\begin{table}[htbp]
\centering
\begin{tabular}{lccc}
\hline
Null model & Below 5\% & Above 95\% & Outside envelope \\
\hline
Shuffle & 11.08\% & 14.55\% & 25.63\% \\
FFT phase-randomized & 46.43\% & 2.91\% & 49.34\% \\
\hline
\end{tabular}
\caption{Pointwise exceedance rates of Bitcoin $L^1$ norm of the persistence landscape relative to surrogate null envelopes.}
\label{tab:null_exceedance}
\end{table}

\section{Application to the S\&P 500 index}
\label{sec:sp500}

\subsection{S\&P 500 data and topological signal}
To examine whether the null-validated topological framework extends beyond cryptocurrency markets, we apply the same pipeline to the S\&P 500 index. This provides a complementary application to a broad and highly liquid equity market benchmark whose volatility structure differs substantially from Bitcoin. Whereas Bitcoin is characterized by larger and more frequent extreme fluctuations, the S\&P 500 exhibits more persistent volatility regimes associated with macroeconomic and equity-market stress. This comparison allows us to assess whether the persistence landscape norm captures market organization in a conventional equity index as well as in a cryptocurrency market.

\begin{figure}
    \centering
    \includegraphics[width=0.9\textwidth]{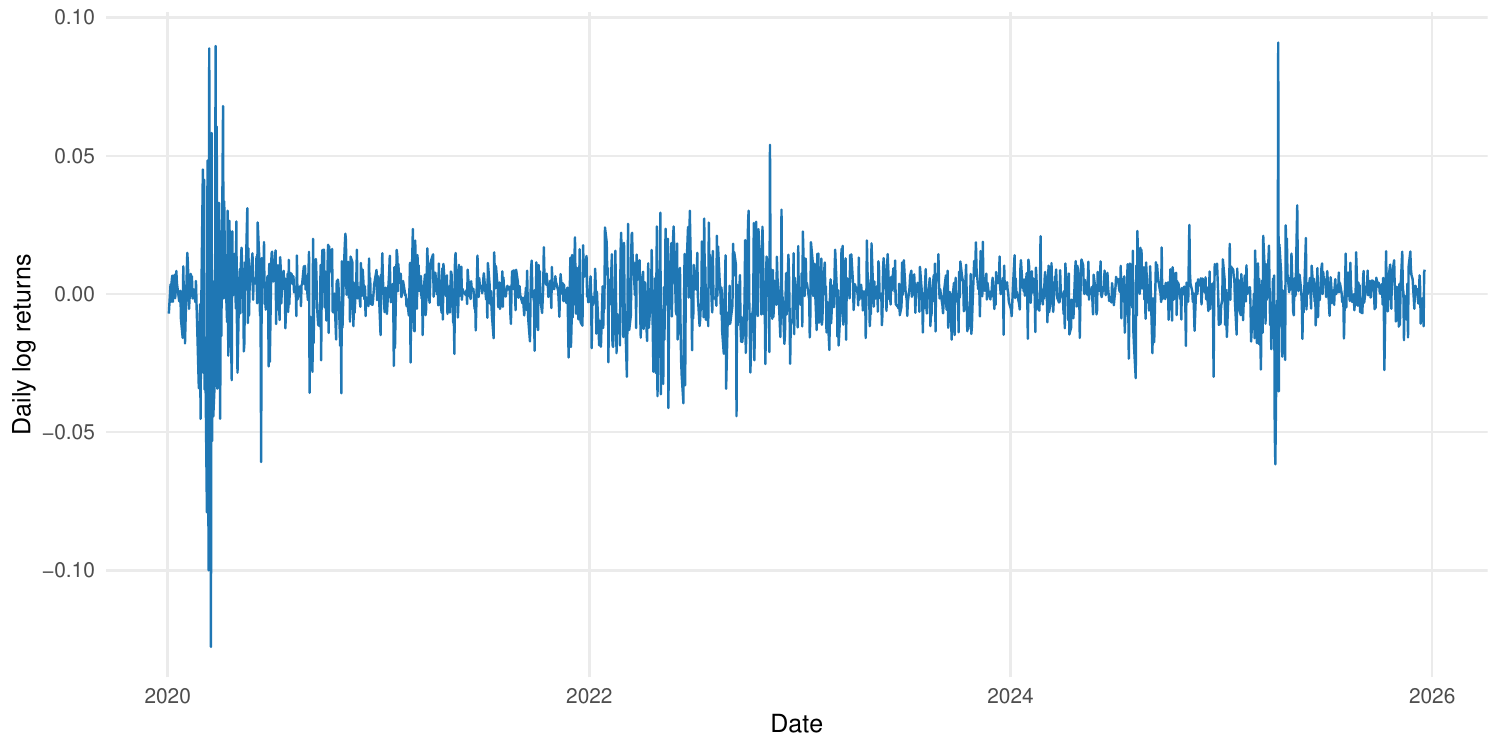}
    \caption{S\&P 500 daily log returns.}
    \label{fig:sp500_returns}
\end{figure}

Daily S\&P 500 closing prices from January 1, 2020 to December 20, 2025 are obtained from Yahoo Finance \cite{Yahoofinance} and used to compute close-to-close trading-day log returns. Figure~\ref{fig:sp500_returns} shows the resulting daily log return series. The largest return fluctuations occur during the early 2020 COVID-19 market shock, with additional volatility episodes visible during 2022 and later localized stress periods.

\begin{figure}
    \centering
    \includegraphics[width=0.9\textwidth]{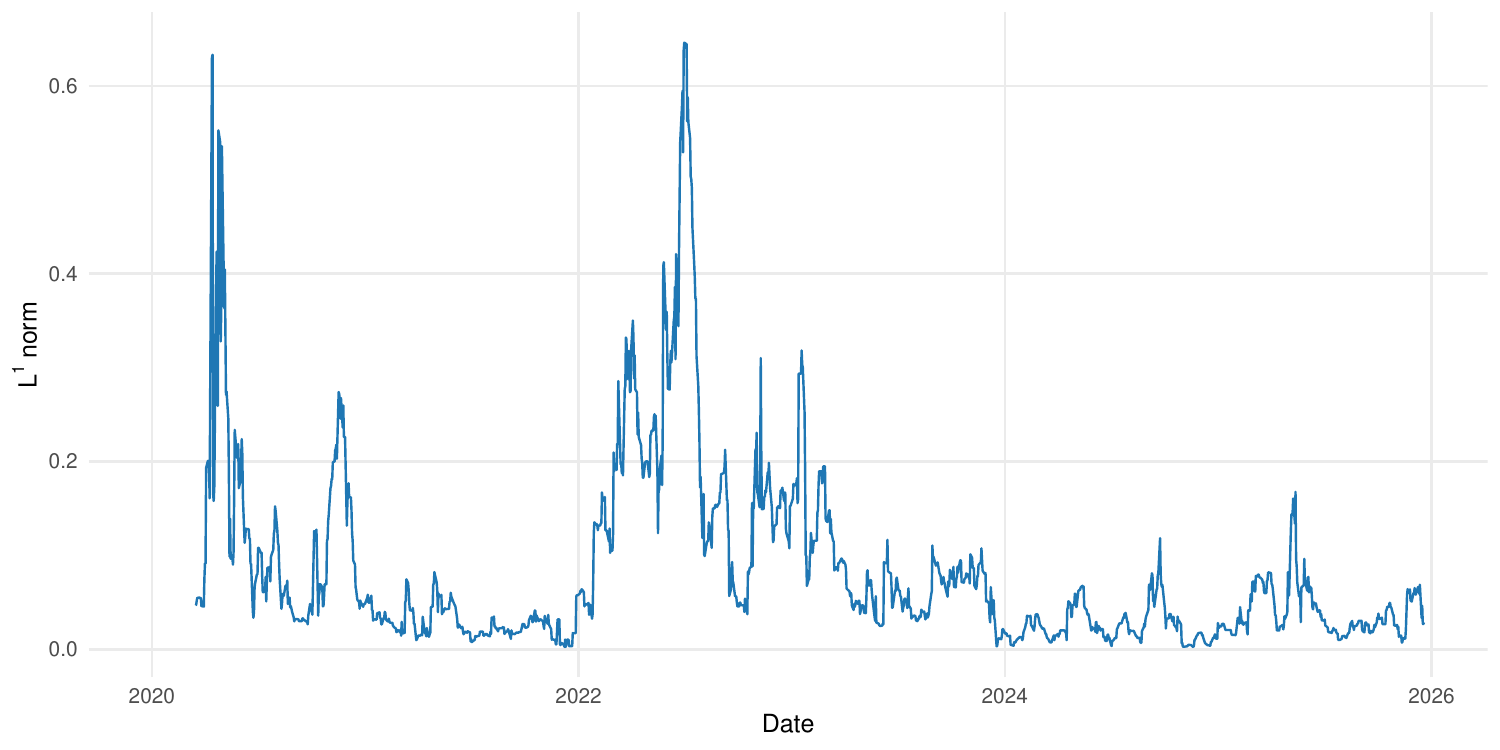}
    \caption{\(L^1\) norm of the persistence landscape computed from sliding-window delay embeddings of standardized S\&P 500 log returns for $w=50, m=4, \tau=2$.}
    \label{fig:sp500_l1}
\end{figure}

For the topological analysis, the S\&P 500 log return series was standardized globally and then analyzed using the same parameters as in the Bitcoin application: embedding dimension $m=4$, delay $\tau=2$, sliding window length $w=50$, one-dimensional persistent homology, and the $L^1$ norm of the persistence landscape. Figure~\ref{fig:sp500_l1} shows the resulting time series of the norm of the persistence landscape.

\subsection{Comparison with stochastic volatility}
We next fit the same stochastic volatility model described in Section~\ref{sec:sv_model} to the S\&P 500 return series. The filtered conditional volatility estimate exhibits a dominant spike during the early 2020 market crash, followed by a lower but persistent elevated-volatility regime during 2022. Compared with Bitcoin, the S\&P 500 filtered volatility is lower in magnitude and more regime-like, which is consistent with the behavior of a broad equity index.

\begin{figure}
    \centering
    \includegraphics[width=0.9\textwidth]{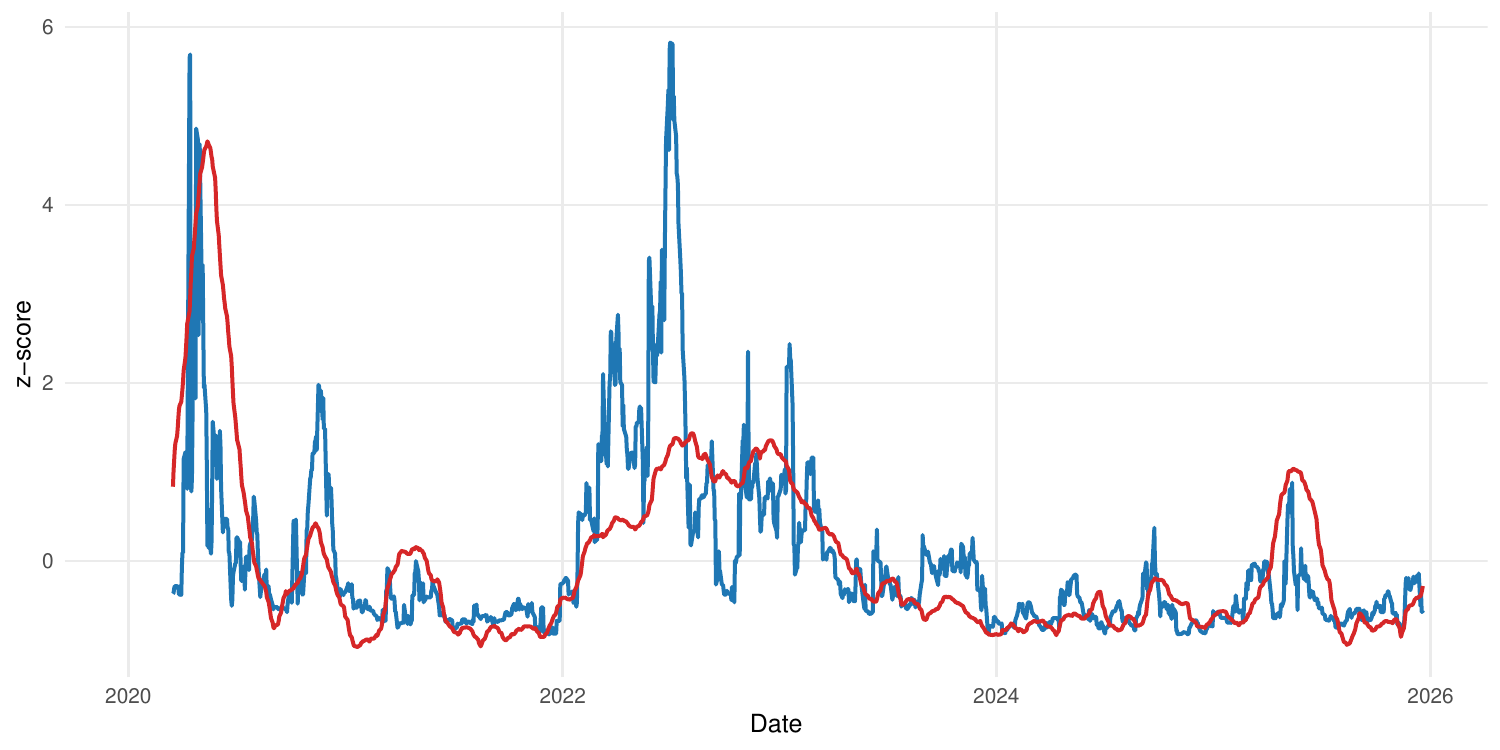}
    \caption{Standardized comparison of the \(L^1\) norm of the persistence landscape (blue) and filtered stochastic volatility (red) for S\&P 500 log returns.}
    \label{fig:sp500_l1_sv_overlay}
\end{figure}

Figure~\ref{fig:sp500_l1_sv_overlay} compares the standardized $L^1$ norm of the persistence landscape with the standardized filtered stochastic volatility estimate. The two quantities co-move strongly during major equity-market stress periods, especially during the COVID-19 crash and the 2022 volatility regime. However, their relationship is not one-to-one. The \(L^1\) signal displays sharper localized bursts, while the stochastic volatility estimate is smoother and more persistent. This indicates that the topological signal is sensitive to volatility, but not identical to the filtered volatility process.

\begin{figure}[htbp]
    \centering
    \includegraphics[width=0.9\textwidth]{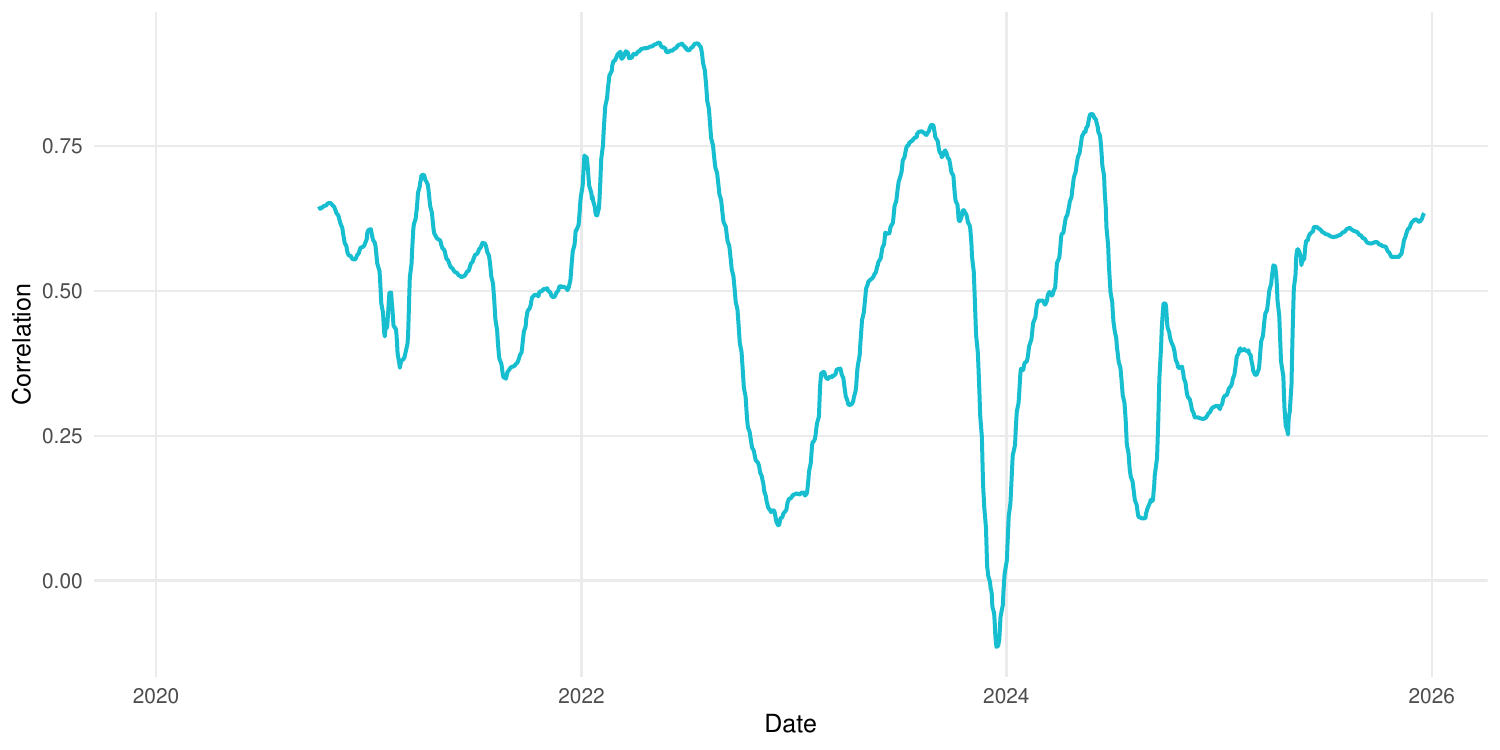}
    \caption{Rolling correlation between the \(L^1\) norm and filtered stochastic volatility for the S\&P 500 using a 180-day window.}
    \label{fig:sp500_rollcorr}
\end{figure}

To quantify the time-varying association between topology and volatility, we compute the rolling correlation between the norm of the persistence landscape and the filtered stochastic volatility estimate using a 180-day window. The result is shown in Figure~\ref{fig:sp500_rollcorr}. The rolling correlation remains mostly positive throughout the sample, indicating a stronger overall coupling between topology and volatility than observed in the Bitcoin application. Nevertheless, the correlation varies substantially over time, with periods of very strong association interrupted by intervals of weaker dependence. In contrast to the Bitcoin case, no single dominant changepoint is detected in the rolling correlation series under the selected penalty. Thus, the S\&P 500 results indicate time-varying coupling without a single sharp transition separating two persistent regimes.

\subsection{Surrogate null validation}
\begin{figure}
    \centering
    \includegraphics[width=0.9\textwidth]{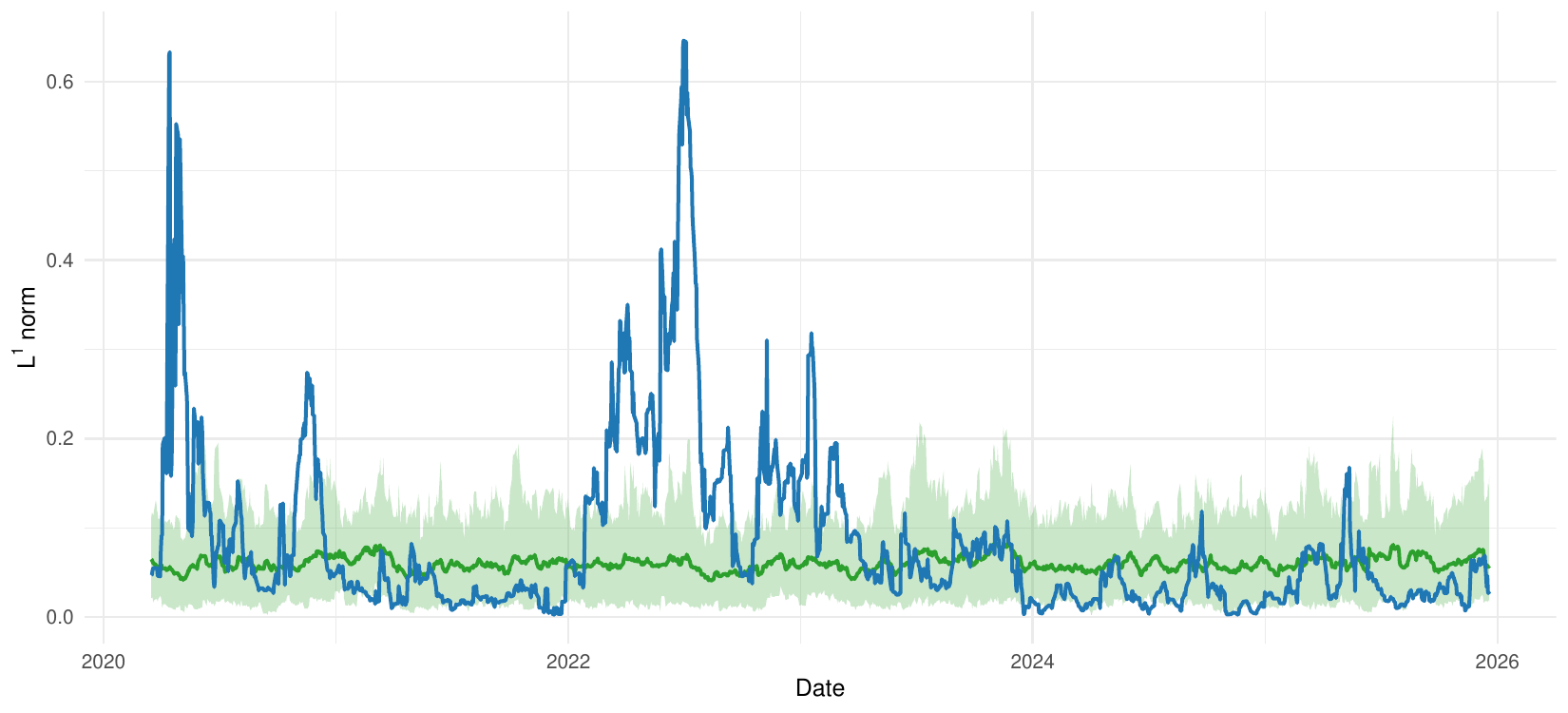}
    \caption{Comparison of the observed $L^1$ norm of the persistence landscape for the S\&P 500 (blue) with pointwise null envelopes obtained from shuffle surrogates.}
    \label{fig:sp500_shuffle_null}
\end{figure}

Finally, we apply the same surrogate-based null validation framework used in the Bitcoin analysis. For the shuffle null, we randomly permute the standardized S\&P 500 return series, thereby preserving the empirical marginal distribution while destroying temporal ordering. Figure~\ref{fig:sp500_shuffle_null} compares the observed \(L^1\) norm with the pointwise null envelope generated from shuffle surrogates. The observed topological signal frequently exceeds the upper shuffle envelope, particularly during early 2020 and 2022. These exceedances indicate that the observed topological structure is not explained by the marginal distribution of returns alone and depends on the temporal organization of the series.

\begin{figure}[htbp]
    \centering
    \includegraphics[width=0.9\textwidth]{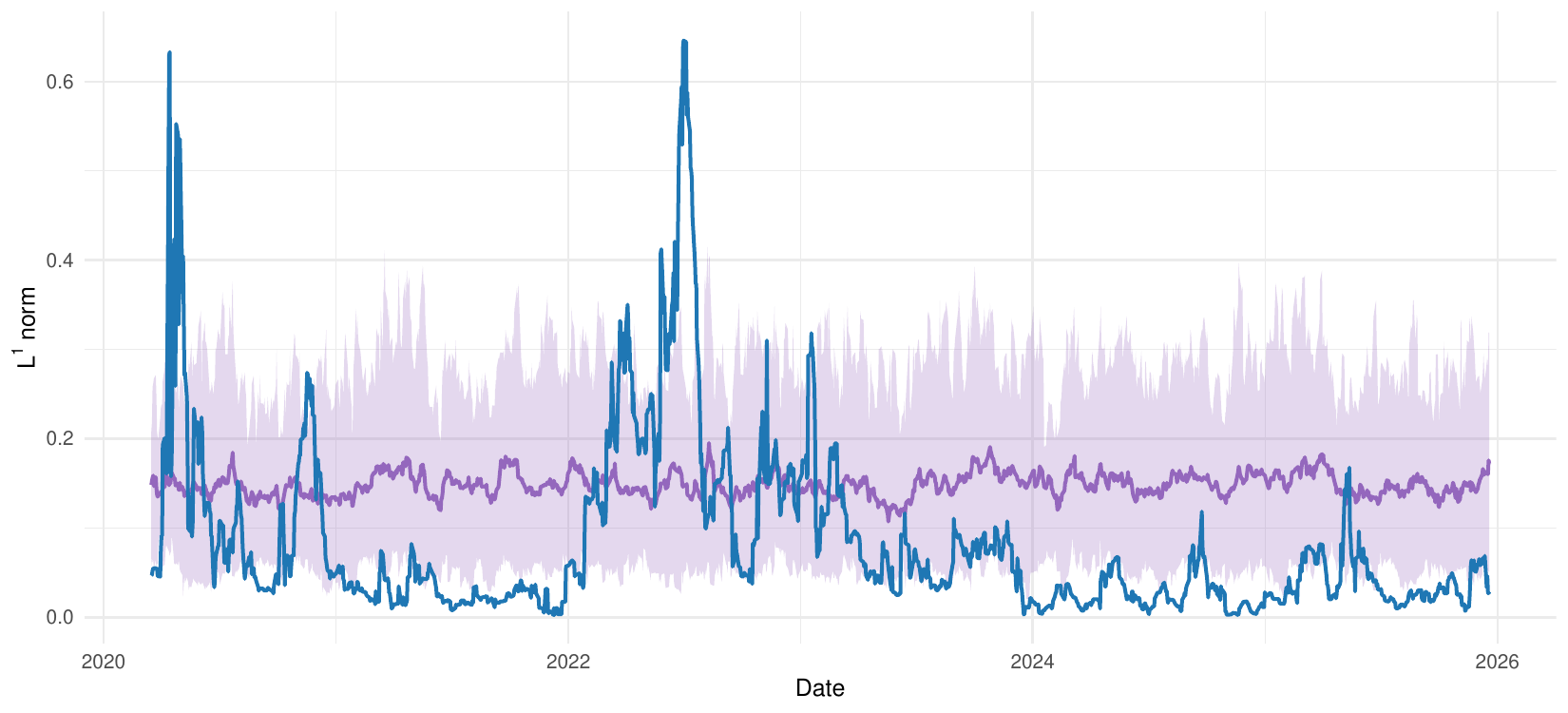}
    \caption{Comparison of the observed $L^1$ norm of the persistence landscape for the S\&P 500 (blue) with pointwise null envelopes generated by FFT phase-randomized surrogates.}
    \label{fig:sp500_fft_null}
\end{figure}

For the FFT phase-randomized null, the Fourier magnitudes are preserved while the phases are randomized, preserving linear second-order structure while disrupting nonlinear and phase-dependent temporal organization. Figure~\ref{fig:sp500_fft_null} shows that the observed \(L^1\) norm also departs substantially from this null. In contrast to the shuffle case, the departures from the FFT null are often below the surrogate envelope, indicating that the observed S\&P 500 dynamics generate less loop-like geometry than many phase-randomized series with the same power spectrum. This behavior suggests that the real temporal organization of S\&P 500 returns constrains the geometry of the delay embedding in a way that is not reproduced by phase randomization.

Table~\ref{tab:sp500_null_exceedance} summarizes the pointwise exceedance rates of the observed $L^1$ norm relative to the surrogate envelopes. The exceedance rates are above the corresponding nominal pointwise levels for both null models, indicating that the S\&P 500 topological signal is not reducible to either marginal return distributions or linear second-order dependence alone.

Overall, the S\&P 500 results show that the null-validated topological signal is not specific to Bitcoin. The norm of the persistence landscape co-moves with stochastic volatility during major equity market stress periods, but its relationship with volatility remains time-varying, and the surrogate analyses indicate departures from null models based on both the marginal distribution and the linear spectrum.

\begin{table}[htbp]
\centering
\begin{tabular}{lccc}
\hline
Null model & Below 5\% & Above 95\% & Outside envelope \\
\hline
Shuffle & 12.06\% & 22.33\% & 34.39\% \\
FFT phase-randomized & 52.65\% & 6.00\% & 58.65\% \\
\hline
\end{tabular}
\caption{Pointwise exceedance rates of the S\&P 500 $L^1$ norm of the persistence landscape relative to surrogate null envelopes.}
\label{tab:sp500_null_exceedance}
\end{table}

\section{Conclusions}
\label{sec:conclusions}

We introduced a null-validated topological framework for quantifying financial market complexity and applied it to Bitcoin daily log returns and the S\&P 500 index as examples of cryptocurrency and broad U.S. equity-market dynamics. The analysis treats the $L^1$ norm of the persistence landscape derived from sliding window delay embeddings as a scalar summary of reconstructed state-space geometry. Across both markets, this quantity is sensitive to periods of market stress, but it is not merely a restatement of volatility. Instead, it provides a complementary descriptor of temporal organization in financial return dynamics.

Comparison with filtered stochastic volatility estimates shows that persistence-landscape norms and volatility encode related but distinct information. During high stress regimes, the two quantities tend to co-move, reflecting the joint presence of large fluctuations and pronounced geometric structure in delay space. However, the strength and stability of this relationship differ across markets. In Bitcoin, the topology-volatility coupling is more unstable and exhibits a clearer regime shift, with periods in which the topological signal persists or fluctuates even as volatility contracts. In the S\&P 500, the coupling is more consistently positive, especially during the COVID-19 crash and the 2022 equity-market stress regime, but the relationship remains time-varying and the topological signal displays localized bursts that are not identical to the smoother filtered volatility process.

For Bitcoin, the residual analysis further shows that temporal structure remains in the topological signal after accounting for stochastic volatility and coarse sentiment variation. This indicates that the norm of the persistence landscape is not fully explained by conventional scale-based measures or by broad sentiment effects. More generally, the rolling correlation analyses for both Bitcoin and the S\&P 500 show that the dependence between topology and volatility is regime-dependent rather than fixed over time.

Surrogate-based null models provide statistical validation of these observations. Rejection of shuffle surrogates demonstrates that the persistence-landscape norm depends on temporal ordering and is not determined solely by the marginal distribution of returns. Departures from FFT phase-randomized surrogates further indicate sensitivity to nonlinear and phase dependent temporal organization beyond linear second order structure. The S\&P 500 application confirms that these surrogate-null departures are not restricted to cryptocurrency data, while also showing that the direction and strength of the departures can differ between the two markets. Thus, the framework detects both common and market-specific forms of temporal organization.

Overall, the results show that nontrivial geometric structure can persist in financial returns even after accounting for volatility, and, in the Bitcoin case, sentiment effects. The topological framework therefore offers a principled approach for detecting regime-dependent organization in financial time series and for distinguishing topological structure arising from marginal distributions, linear dependence, and higher-order temporal dynamics.

Several directions remain for future work. A systematic cross-asset study involving additional equity indices, individual equities, fixed income instruments, commodities, and foreign exchange markets would help clarify how topology-volatility coupling varies across market structures and time scales. From a methodological perspective, future work may incorporate more restrictive surrogate constructions, such as iterated amplitude adjusted Fourier transform surrogates, to further disentangle nonlinear temporal structure from linear dependence and marginal distribution effects. It would also be useful to investigate the mechanisms driving pronounced peaks in persistence landscape norms and to study whether these topological signatures have predictive value for regime shifts, market stress, or changes in volatility dynamics.

\section*{Data availability}
Bitcoin and S\&P 500 price data used in this study are publicly available from Yahoo Finance. The Crypto Fear \& Greed Index data used in the Bitcoin sentiment analysis are publicly available from alternative.me.

\section*{Declaration of competing interest}
The author declares that there are no known competing financial interests or personal relationships that could have appeared to influence the work reported in this paper.

\section*{Acknowledgement}
The author thanks Marian Gidea for helpful discussions.

\FloatBarrier

\bibliographystyle{unsrt}  
\bibliography{tda_vol}

\end{document}